\documentstyle[prl,aps,twocolumn,epsf,floats]{revtex} 
\begin{document}

\wideabs{  
\draft 
\title{ Strong, Ultra-narrow Peaks of Longitudinal and Hall 
 Resistances in the Regime of Breakdown of the Quantum Hall Effect}

\author{ A. M. Song and P. Omling }
\address{ Solid State Physics/Nanometer 
 Structure Consortum, Lund University, Box 118, S-221 00 
         Lund, Sweden}
\date{\today}
\maketitle
\begin{abstract} 
With unusually slow and high-resolution sweeps of magnetic field, 
strong, ultra-narrow (width down to $100\ {\rm \mu T}$) 
resistance peaks are 
observed in the regime of breakdown of the quantum Hall effect. 
The peaks are dependent on 
the directions and even the history of magnetic field sweeps, 
indicating the involvement of a very slow physical process. 
Such a process and the sharp peaks are, however, not predicted 
by existing theories. 
We also find a clear connection between the resistance peaks 
and  nuclear spin polarization. 
\end{abstract}

} 

The integer quantum Hall effect (QHE) is a most remarkable phenomenon of 
two dimensional electron system (2DES), in which 
the Hall resistance  is 
quantized to $h/ie^2$  while the longitudinal 
resistance nearly vanishes ($h$ is Planck's constant, $e$ 
the electron charge, and $i$ an integer)\cite{1}.
To employ the QHE for the resistance standard, it is desirable to 
apply a high current through a Hall bar. 
However, it was early discovered that the QHE breaks down 
if the current reaches a critical value, $I_{c}$\cite{2,3}. 
Extensive investigations were thereafter
performed to study the origin of the 
breakdown\cite{4,5,6,7,8,9,10,11,12,13,14,15}. 
So far, most studies have focused on factors that influence the 
critical current around, in particular, even filling factors. 
A number of models have been proposed, such as inter-Landau-level 
scattering\cite{4,7} and the superheating process\cite{2,5}. 
However, the exact mechanism responsible for the  
breakdown is still under debate. 

Here, we report on the measurement of  
the {\em differential} 
longitudinal and  Hall resistances $R_{xx}$ and
$R_{xy}$ (the derivative of voltage with 
respect to the total applied current) 
at high injected currents close to $I_{c}$. 
With unusually slow, high-resolution sweeps of magnetic field $B$, 
ultra-narrow $R_{xx}$ peaks (width down to $100\ {\rm \mu T}$)
are observed. The peak values exceed the resistances of the surrounding 
magnetic fields by a factor 36. While no substantial change in 
$R_{xy}$ is noticed 
around the odd filling factor $\nu=3$, strong, sharp peaks are 
also shown on the $R_{xy}$ curves for $\nu=2$ 
and 4. We find the peaks to be sensitively dependent on 
the directions and even the history of the $B$ sweeps. 
This indicates that a physical process with a very large time 
constant is involved, which is orders of magnitude longer than 
that may be predicted by 
the existing models for the QHE breakdown.  
While many {\em disordered} electronic systems have recently been 
found to exhibit very 
slow relaxations\cite{16},  to our knowledge, 
the unusually slow physical process to be reported here has never been 
observed in the integer QHE regime.  
Interestingly, some of the aspects of our experimental observations 
are similar to the  discovered anomalous
resistance peaks in the {\em fractional} QHE
regime\cite{17}, while some other aspects are apparently different. 
We will also show that the sharp 
resistance peaks are influenced by the nuclear spin flips. 
Furthermore, we present a model, which qualitatively explains the 
different aspects of our observations. 

We use two GaAs/AlGaAs modulation-doped 
heterostructures (wafer I and wafer II) with 
carrier densities of  $n_{s}=3.7$ and 
$3.5\ \times\ 10^{15}\ {\rm m^{-2}}$ and 
mobilities of  $\mu=59$ and $130\ {\rm m^2/Vs}$ at 0.3\ K, 
respectively. 
A modulation-doped ${\rm In}_{0.75}{\rm Ga}_{0.25}$As/InP 
structure ($n_{s}=2.8\ \times\ 10^{15}\ {\rm m^{-2}}$, 
$\mu=22\ {\rm m^2/Vs}$) is also studied.  
For all these wafers, $I_{c}$ is found to scale linearly 
with the device width. 
The experiments are performed in a $^{3}$He 
refrigerator at 0.3\ K. Hall devices with different 
widths (from 43 to
$200\ \mu{\rm m}$) and different geometries are
investigated using a standard lock-in technique with a  
frequency of 17\ Hz. Together with 
a 5\ nA ac current, large dc currents, $I_{dc}$, 
are sent through the sample 
to drive the 2DES close to 
the regime of breakdown of the QHE. 
Qualitatively similar behavior is observed in all the samples  
fabricated from different material systems. 
We report here on measurements performed on a 
Hall bar made from wafer I.  

\begin {figure}
\centerline{
\epsfxsize=\columnwidth
\epsffile{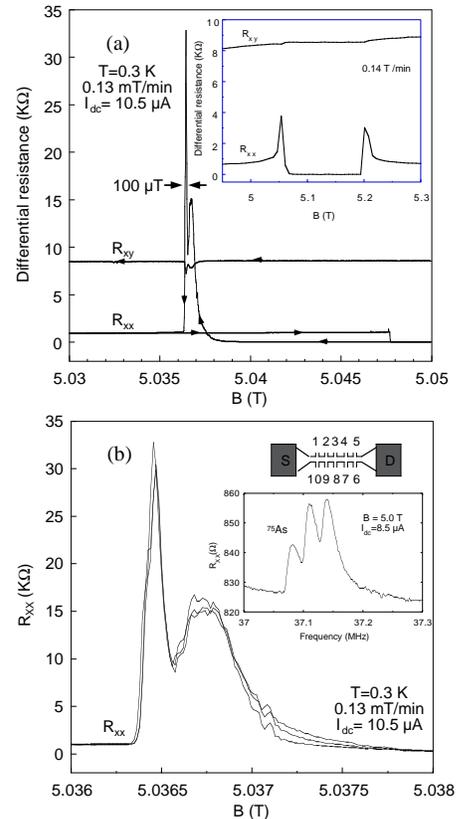} }
\caption { 
(a) $R_{xx}$ and $R_{xy}$ as a function of $B$ around $\nu=3$ at  
$I_{dc}=10.5\ \mu{\rm A}$. The sweep speeds are $0.13\ {\rm mT/min}$ 
(main curves) and $0.14\ {\rm T/min}$ (inset). 
(b)  $R_{xx}$ curves of a downward sweep from 
different segments of the Hall bar, which is shown in the upper inset.  
The lower inset shows an NMR 
resonance spectrum of  ${\rm ^{75}As}$. 
The applied dc current is   
$8.5\ \mu$A, with which a sharp $R_{xx}$ peak 
is detected at 5.0\ T. By fixing $B$ at 5.0\ T, however, 
we find that $R_{xx}$ slowly decreases with time and after about 15 
minutes becomes stabilized at about 825 $\Omega$. By applying 
rf signal to a small coil around the sample, a clear 
threefold splitting  is observed. } 
\label{f1}
\end {figure}

The inset of Fig.\ \ref{f1}(a) shows the curves of 
differential resistances $R_{xx}$ and
$R_{xy}$ as a function of $B$ around  
$\nu=3$ and at a dc current close to, but below, the 
critical current $I_{c}=11\ \mu{\rm A}$. The Hall bar has 
a width of $43\ \mu{\rm m}$ and 
five pairs of voltage probes, as is schematically
shown in the inset of Fig.\ \ref{f1}(b). 
The $B$ sweep is at a ``normal''  speed 
of 0.14\ T/min and the curves are ``as expected'', i.e.   
$R_{xx}$ nearly vanishes and $R_{xy}=h/3e^2$ within a  
$B$ range (i.e. the dissipationless regime) 
that is narrower than that at $I_{dc}=0$. 
However, by reducing the sweep speed and increasing 
the magnetic field resolution, the two
$R_{xx}$ peaks at the left and right edges of the dissipationless regime
become successively higher  
and narrower. Furthermore, the curves of the upward and downward 
sweeps become increasingly different. 
Figure\ \ref{f1}(a) shows the differential resistances around the left edge 
of the dissipationless regime at a sweep speed of 
$0.13\ {\rm mT/min}$ and a sweep step of 0.000015\ T, which is the 
resolution of our magnet system. 
The arrows on the curves indicate the sweep directions. 
While the downward sweeps show only very small changes in $R_{xy}$, 
strong peaks are observed on the $R_{xx}$ curve. The 
narrower peak has a full width at half maximum (FWHM) of only 
$100\ {\rm \mu T}$. The resistance value at the peak is almost four times 
as high as the value at the Hall 
plateau and about 36 times higher than the $R_{xx}$ value on the 
lower $B$ side. For lower magnetic fields, $R_{xx}$ is found 
to remain virtually constant\cite{18}. When sweeping upwards from 5.03\ T, 
however, $R_{xx}$ remains at this constant value (no peak structures) 
until it suddenly drops to zero at about 5.048\ T. 
The behavior is thus totally 
different from the hysteresis effect of the breakdown of the 
QHE\cite{2,3} where only a shift in the magnetic field 
position is observed. 

We have simultaneously measured $R_{xx}$ using different 
segments of the Hall bar. Figure\ \ref{f1}(b) shows the results of a 
downward sweep within 2\ mT, using probes 1 and 2, 2 and 3, and 4 and 5. 
We obtain almost identically strong, narrow resistance peaks from 
different parts of the Hall bar. For instance, it can be seen
in Fig.\ \ref{f1}(b) that the 
peak on the higher field side has a fine structure, which can be seen  
on all the three traces. 
This rules out the possibility 
that our observations are due to local breakdown 
induced by  inhomogeneities of the sample.  
Further studies of the fine structure, however, require a magnet 
system with a better resolution.
The behavior of $R_{xx}$ and $R_{xy}$ 
at the higher $B$ edge of the dissipationless regime is 
very similar to that shown in Fig.\ \ref{f1}. 
There, an upward sweep results in sharp  $R_{xx}$
peaks while a downward sweep shows only a sudden drop to zero. 

\begin {figure}
\centerline{
\epsfxsize=\columnwidth
\epsffile{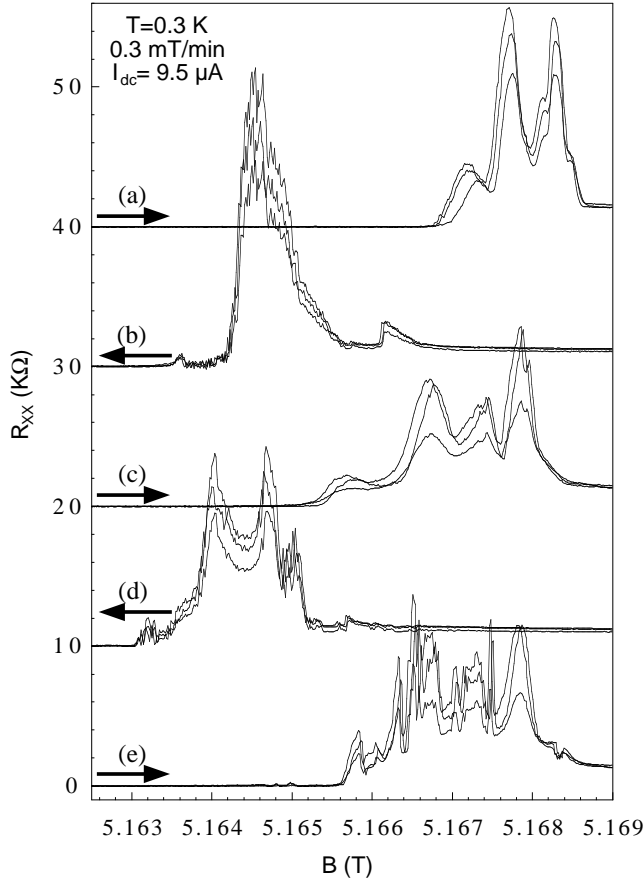} }
\caption { 
$R_{xx}$ traces obtained from different 
segments of the Hall bar at  $I_{dc}=9.5\ \mu$A and  
around the higher $B$ edge of the dissipationless regime of $\nu=3$. 
As it is a different cooling cycle, the peak position is lower than 
that shown in the inset of Fig.\ \ref{f1}(a) although  
it should have been shifted towards a higher field because $I_{dc}$ is 
lower. Five successive sweeps [(a)--(e), offset by 
$10\ {\rm k\Omega}$ for clarity] are made 
back and forth at a speed of 0.3\ mT/min. 
The arrows indicate the sweep directions. 
}
\label{f2}
\end {figure}

If $I_{dc}$ is decreased, the height of 
the $R_{xx}$ peaks is reduced, while the width increases. 
Furthermore, there is less difference in the $R_{xx}$ curves 
between upward and downward sweeps. 
Figure\ \ref{f2} shows the $R_{xx}$ traces obtained from different 
segments of the Hall bar at a lower current, $I_{dc}=9.5\ \mu$A. 
The $B$ range corresponds to the right edge of the 
dissipationless regime around $\nu =3$. 
Five successive sweeps [Figs. (a)--(e)] are made 
back and forth between $5.160\ $T 
and $5.175\ $T with a speed of 0.3\ mT/min. 
The curves are plotted only in the range between 
5.1625\ T and 5.1690\ T for clarity. 
The change in the peak position of about 3\ mT with sweep direction
is most likely due to hysteresis of the magnet system. 
Although each sweep takes about one hour, the 
peak structure changes gradually, 
indicating the involvement of a very slow physical process.
We have noticed the following points. First, 
curves obtained in the same sweep direction,
such as Figs.\ (a), (c), and (e) or Figs. (b) and (d), are  
similar. Second, the greater the number of sweeps made, 
the less the difference in the $R_{xx}$ curves 
between upward and downward sweeps. This can already be seen from the 
increased similarity between   
Figs.\ (d) and (e), and is more clear in later sweeps (not 
shown here). Third, an increasing number of peaks and fine structures are 
obtained when more sweeps are made. 
This rules out any trivial heating effects, as heating is 
expected to smear out fine structures. 

\begin {figure}
\centerline{
\epsfxsize=9cm  
\epsffile{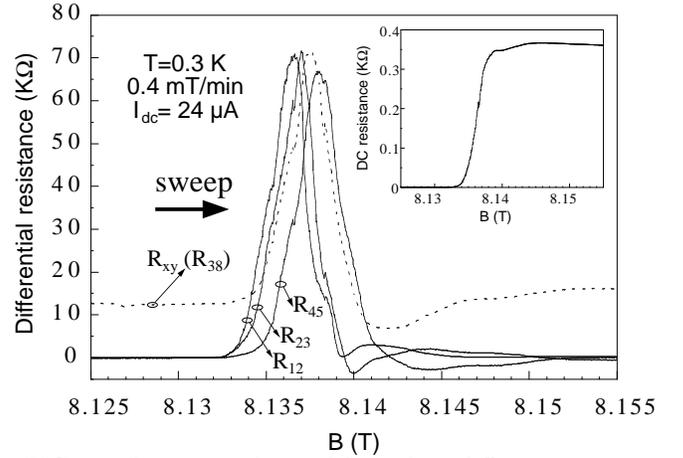} }
\caption { 
$R_{xx}$ traces obtained from three different 
segments of the Hall bar and one $R_{xy}$ curve, 
at $I_{dc}=24\ \mu$A and around the higher  $B$
 edge of the dissipationless regime of $\nu=2$.
The inset shows a simultaneous measurement of the dc longitudinal 
resistance. 
}
\label{f3}
\end {figure}

Although strong peaks are observed on the $R_{xx}$ curves, the Hall 
resistance around  $\nu =3$
shows only small changes 
as can be seen in 
Fig.\ \ref{f1}(a). The behavior of $R_{xy}$ around the even filling 
factors $\nu =2$ and 4
is, however, totally different. 
This suggests that the phenomenon is connected 
with the spin of the 2DES.
Figure\ \ref{f3} shows three $R_{xx}$ traces taken from different 
segments of the Hall bar and one $R_{xy}$ curve around $\nu =2$.  
The dc current is $24\ \mu$A,  
which is about $I_{c}/2$ at this filling factor\cite {19}. 
The $B$ range is centered at the right edge of the 
dissipationless regime. 
In contrast to the results for odd filling factors [see Fig.\ 
\ref{f1}(a)], an equally  
strong, narrow peak (FWHM below 3\ mT)
forms on the $R_{xy}$ trace as on the $R_{xx}$ traces. 
The peak value
is more than five times higher than the Hall plateau $h/2e^2$. 

It can be observed that $R_{xx}$  becomes negative on the higher $B$ 
side of the peaks in Fig.\ \ref{f3}. A dc measurement of the 
longitudinal resistance is shown in the inset. 
Obviously, dc resistances can be 
quite different from differential resistances in the nonlinear regime. 
This is the reason why no anomalous behavior is observed in the dc 
measurement at $8.1354\ $T where sharp peaks form on the differential 
resistance curves. 
In fact, we do not see any unusual behavior of the 
dc resistance
at other $B$ values.

The general features reported here 
are observed at all filling 
factors at sufficiently high magnetic fields and  
in all the Hall bars and wafers studied. Thus, the 
above phenomena seem to be
general in 2DES.  The fine structures are, however, very difficult to 
fully reproduce in different samples. This is, at least in part, due to 
the fact that the fine structures are extremely sensitive to 
 the exact $I_{dc}$ used, sweep speed, starting point of sweeps, 
history, etc. 

While many disordered electronic systems are characterized by very 
slow relaxations\cite{16},  to our knowledge, 
the above unusually slow physical process has never been 
observed in the integer QHE regime.
It is orders of magnitude slower than 
the time scale of the instabilities  
in the regime of the QHE breakdown \cite{3,8,9}. 
The existing models for the breakdown of the QHE, 
such as inter-Landau-level 
scattering\cite{4,7} and electron superheating\cite{2,5}, 
do not predict any physical process
with a time constant larger than microseconds. 
Interestingly, we have noticed that 
certain aspects of our observations, 
such as the long time constant, strong $R_{xx}$ peaks, and current 
dependence, are similar to the recently discovered anomalous
resistance peaks in the {\em fractional} QHE regime at
$\nu=\frac {2}{3}$ and $\frac {3}{5}$\cite{17}. 
However, some other aspects are different, such as 
the existence of fine structures, strong $R_{xy}$ peaks, 
the much sharper peaks (more than three orders of magnitude narrower), 
etc. Very recently, the peaks 
observed in Ref. \cite{17} were found to be influenced by the  
nuclear spin polarization\cite{20}. 
We have also performed 
nuclear magnetic resonance (NMR)
experiments on ${\rm ^{75}As}$, ${\rm 
^{69}Ga}$, and ${\rm ^{71}Ga}$. A typical result for ${\rm ^{75}As}$
is shown by the lower inset in Fig.\ \ref{f1}(b). 
The splitting of the line is, however, 
threefold that is different 
from the fourfold splitting observed in Ref. \cite{20}. 
Furthermore, we observe
resonance peaks rather than dips as in Ref. \cite{20}. 
While the above NMR response is strong in the GaAs/AlGaAs samples, 
so far, no clear observation has been obtained in our InGaAs/InP samples. 
One reason might be the comparatively low mobility of those samples.

In the following, we present a  
model, which qualitatively explains the different aspects of our 
observations. In the $B$ range of 
a dissipationless regime, the bulk of the Hall bar is actually insulating. 
In the single-particle picture, 
if $B$ is sufficiently high, each Landau level is split into
two well-separated, spin-polarized levels with a degeneracy 
proportional to $B$. Therefore, a change in  $B$  will induce a 
redistribution of the electrons in the bulk of 
the Hall bar (denoted ``bulk electrons'')
among the Landau levels, i.e. 
some electrons need to {\em have their energies changed 
and their spins flipped} in order to achieve equilibrium. 
However, as the bulk electrons have no effective 
interaction with the electrons at the edge nor with electron 
reservoirs (the ohmic contacts) in the dissipationless regime, 
the  scatterings required to flip the spins and change the energies 
are virtually absent. The redistribution among the {\em single-particle} 
Landau levels is thus not possible. 
This means that the bulk electrons can be far from the
``normal equilibrium'' (the equilibrated distribution 
among the single-particle Landau levels)
inside the dissipationless regime if $B$ is changed. 
To the best of our knowledge,  no study has been carried out
on how these electrons redistribute in the energy and spin space 
in such a ``nonequilibrium'' situation.  
As it is not possible for the bulk electrons to redistribute 
among the single-particle 
Landau levels, effects such as electron-electron interactions must 
take place. We speculate that the real distribution maintains some order, 
which means that the electrons might rearrange to form ``mini-gaps'' and 
``mini-bands'' in the energy and spin distribution. 

When the 2DES  
starts to enter the dissipation regime where the bulk-edge
interactions are still considerably weak, we expect  
the electrons in the mini-bands to be affected. 
Each time a mini-band starts to participate in the scattering 
process, a differential resistance peak is  observed. In this 
picture, the multiple resistance peaks and fine structure
reflect the mini-band structure of the nonequilibrium distribution of the 
bulk electrons. 
One may speculate that similar nonequilibrium distribution
also forms in the {\em fractional} QHE 
regime\cite{21}, which might as well give rise to resistance 
peaks.  If a strong current is 
applied to the Hall bar, the large Hall electric field will 
substantially enhance the interaction between the electrons at the edge 
and those in the bulk, and therefore give rise to much stronger and sharper 
resistance peaks, in agreement with our experimental observations. 
Note that the ranges of $B$ in which the resistance peaks and 
fine structures are observed are only {\em slightly } 
away from the dissipationless regime. Therefore, the scattering
between electrons in the bulk and electrons at the edge is expected 
to be rather weak. In addition, since the bulk area of a Hall bar is 
fairly large, the time constant of the equilibration can be very long, 
which explains 
the slow physical process indicated especially in Fig.\ \ref{f2}. 
The details of the distribution of nonequilibrium electrons, and 
thereby the mini-bands, depend on the initial $B$ position, 
the sweep direction, and the sweep speed. This 
thus explains the 
observed strong dependence of the resistance peaks and fine structures on 
the experimental history.

The observed NMR resistance peaks shown in the inset of 
Fig.\ \ref{f1}(b) also supports our model. 
Via the hyperfine interaction an 
electron spin can flip with a simultaneous flop of a nuclear spin, 
which can be induced by, for example, applying NMR rf signals\cite{22}. 
Because in our model the 
lack of electron spin-flip scatterings is the reason for the 
nonequilibrium distribution of bulk electrons, the 
additional electron spin-flip scattering induced by the NMR signals 
will reduce the degree of nonequilibrium
distribution. This leads to an 
increased scattering probability from edge to bulk, which is 
detected as an increase of the resistance, as shown 
in the lower inset of Fig.\ \ref{f1}(b).
The threefold splitting is most likely caused by the electric 
quadrupole interaction, 
which is possible in our sample where large 
electric field gradients are expected. 
This connection to nuclear spins is in line with earlier observations 
of the importance of nuclear spin polarization in experiments on 2DES. 
The dynamical nuclear polarization has been observed as Overhauser 
shifts in electrically detected spin resonance experiments 
\cite{23} and in e.g. the 
time dependency of current-voltage characteristics in transport 
experiments in which spin polarized electrons were injected 
\cite{24,25,26}.  Also, our results are, although 
performed in a different physical regime, similar to the recent 
findings in Ref. 20. This may imply that a similar scattering mechanism 
might be involved in the two different regimes.

To conclude, unexpected strong, ultra-narrow resistance
peaks and fine structures have been observed in the regime of breakdown 
of the QHE. The studies reveal the involvement of a
very slow physical process, which is not predicted by existing 
models. 
We also show a clear connection between the sharp peaks 
and the nuclear spin polarization. 
Furthermore, 
we have presented a model that emphasizes the important role of 
the nonequilibrium distribution of bulk electrons and 
qualitatively explains the observed phenomena. 

We acknowledge useful discussions with H. Q. Xu and technical 
support by A. Svensson and H. Persson. This work was 
supported by the Swedish Natural Science Research Council.

\end {document}